\documentclass[aps,prc,twocolumn]{revtex4}

\usepackage{graphicx}
\usepackage{amsmath}
\usepackage{dcolumn} 
\usepackage{amsfonts}
\usepackage{amssymb}
\usepackage{color}

\newcommand{\be}{\begin{equation}}
\newcommand{\ee}{\end{equation}}
\newcommand{\ba}{\begin{array}}
\newcommand{\ea}{\end{array}}
\newcommand{\bea}{\begin{eqnarray}}
\newcommand{\eea}{\end{eqnarray}}
\def\hf{\textstyle{1\over2}}

\begin{document}


\title{An equations-of-motion approach to quantum  mechanics: \\ application to a model phase
transition}


\author{S.Y.\ Ho$^\dag$,  G.~Rosensteel,$^\ddag$ and D.J.~Rowe,$^\dag$}
\affiliation{$^\dag$Department of Physics, University of Toronto,
Toronto, Ontario M5S 1A7, Canada \\
$^\ddag$Department of Physics, Tulane University,
New Orleans, LA 70118 USA}

\date{Oct.\ 10, 2006}  

\begin{abstract}
We present a generalized equations-of-motion method that efficiently calculates energy 
spectra and matrix elements for algebraic models. The method is applied to a 5-dimensional
quartic oscillator that exhibits a quantum phase transition between vibrational and rotational
phases. 
For certain parameters, $10\times 10$ matrices give better results than
obtained by diagonalising 1000 by 1000 matrices.


\end{abstract}


\maketitle

Equations-of-motion methods \cite{Sawada,Kerman,KK,RowRMP} offer an alternative to 
diagonalization of a Hamiltonian to determine the properties of a quantal system.
We consider systems with a Hamiltonian  expressed in terms of a Lie algebra of observables,
$\frak{g}$. 
For a finite-dimensional irreducible representation (finite irrep) of $\frak{g}$, both
approaches  give the same results. However, they become inequivalent when approximations are
necessary. For example, diagonalizing the Hamiltonian for a system of $n$ coupled harmonic
oscillators, in a basis of uncoupled harmonic oscillator states, gives results that become
precise (to within computational errors) in an infinite limit whereas an
equations-of-motion approach, as given by the standard random phase approximation (RPA), is
already precise in a $2n$-dimensional space. 

Problems of interest are ones for which an expansion of the
low-energy eigenstates of the Hamiltonian, in a given ordered basis for the Hilbert space,
converges slowly.
For example, the low-energy states of strongly-deformed rotational nuclei 
are dominated by components from  higher shells in contrast to the low-energy states of a
near-spherical nucleus which may be dominated by the valence-shell states of a spherical
harmonic-oscillator shell-model basis \cite{NShapes}.  
In such a situation, diagonalization of the Hamiltonian in a spherical shell-model basis is
unlikely to give reliable results.

The proposed  approach avoids the preliminary
stage of defining basis states and proceeds directly to the determination of a matrix
representation of the algebra of observables in which the Hamiltonian is diagonal.
The approach has its origins in three previous developments: 
a variational technique \cite{Ros,RR} for computing the irreps
of potentially difficult Lie algebras; the double-commutator equations-of-motion
formalism \cite{RowRMP}; and the equations-of-motion method of Kerman and
Klein \cite{KK,LiK}.
Thus, we refer to it as the RRKK equations-of-motion method.

Consider a Hamiltonian $\hat H$ that is a polynomial in the elements  $\{\hat X_\nu\}$ of a
Lie algebra $\frak{g}$ of observables with commutation relations
$[\hat X_\mu, \hat X_\nu]  = \sum_\sigma C^\sigma_{\mu\nu}\,\hat X_\sigma$.
We refer to $\frak{g}$ as a spectrum generating algebra (SGA),
For each $\hat X_\nu\in \frak{g}$, define $\hat {\cal H}_\nu := [\hat H,\hat
X_\nu ]$. The objective is to determine a unitary irrep in which each
observable   $\hat X_\nu$ is represented  by a matrix $X(\nu)$, with elements
\be X_{ij}(\nu) := \langle i |\hat X_\nu |j\rangle, \ee
to be determined along with energy differences $\{E_i-E_j\}$, such that the sets of
equations
\bea 
&& f_{ij}(\mu,\nu) :=\langle i | [\hat X_\mu, \hat X_\nu] |j\rangle - \sum_\sigma
C^\sigma_{\mu\nu} \langle i|\hat X_\sigma |j\rangle =0, \quad\label{eq:1} \\
&&f_{ij}(\nu) :=\langle i | {\cal H}_\nu (\hat X) |j\rangle - (E_i-E_j)\langle i|\hat
X_\nu |j\rangle =0, \quad\label{eq:2}
\eea
are satisfied.
In general, it will also be necessary to include additional equations  to ensure that the
representation is the one desired. 
For example, if the Lie algebra has
Casimir invariants, $\hat C_n(\frak{g})$, equations can be included to require that they
are represented by the appropriate multiples of the unit matrix, $c_nI$.

Let $F$, a so-called \emph{objective function},   be defined for a finite irrep of
dimension
$N$ as the sum of squares
\bea F &=&\sum_{i,j}^N \Big[\sum_{\mu\nu} \big|f_{ij}(\mu,\nu)\big|^2 + 
\sum_{\nu} \big|f_{ij}(\nu)\big|^2 \cr 
&&+ \sum_n\big|\langle i |\hat C_n(\frak{g}) - c_n |j\rangle \big|^2\Big] . \eea
$F$ cannot be negative and can only vanish when a precise
solution to the system of equations has been obtained. Thus, for finite  irreps,
precise solutions to the above equations are obtained by minimization of $F$ as a
function of the unknown matrix elements of the observables and the energy differences.
If needed, the Hamiltonian matrix can also be  evaluated as a polynomial in
the $\{X(\nu)\}$ matrices to determine the ground-state energy.

The challenge is to obtain accurate solutions for finite
submatrices of the observables, corresponding to a subset of lowest-energy eigenstates,  when
the irrep is  infinite.  
Recall that a differential equation defined over the positive  half of the real line, $0
<r <\infty$, can be solved precisely over a finite interval $0< r\leq R$, if one knows the
boundary conditions at $R$.
Similarly, when infinite dimensional matrices $\{X(\mu)\}$ are truncated to finite
dimensions, their outer rows and columns provide boundaries and their entries can be
adjusted to give accurate results for the  matrices they enclose.

The concern with boundary conditions in the equations-of-motion approach arises because,
the equations of motion for a subset of matrix elements $\{X_{ij}(\nu); 1\leq i,j\leq
N\}$, involve the commutation
relations
\be f_{ij}(\mu,\nu) = \sum_{k=1}^{\infty} \big[
X_{ik}(\mu)X_{kj}(\nu)  -X_{ik}(\nu)X_{kj}(\mu) \big] 
\label{eq:extras}
\ee
and, hence, $X_{ik}(\mu)$ matrix elements with $k>N$.

This concern can be resolved as follows:
because the matrix elements connecting lowest-energy
states are of most interest, we apply a weighting factor $w_{ij} = 1/[(i+1)(j+1)]$ to the
expressions that should vanish for an exact solution; i.e., redefine the objective function to
be minimized as
\bea F' &=&\sum_{i,j}^N w^2_{ij}\Big[\sum_{\mu\nu} \big|f_{ij}(\mu,\nu)\big|^2 + 
\sum_{\nu} \big|f_{ij}(\nu)\big|^2 \cr 
&&+ \sum_n \big|\langle i | C_n(\frak{g}) - c_n |j\rangle \big|^2\Big] . \label{OFI}\eea
We also make the approximation
\be f_{ij}(\mu,\nu) \approx \sum_{k=1}^{N+1} \big[
X_{ik}(\mu)X_{kj}(\nu)  -X_{ik}(\nu)X_{kj}(\mu) \big] 
\label{eq:N+1}
\ee
and corresponding approximations for the evaluation of $f_{ij}(\nu)$ and the matrix elements
of Casimir invariants.
 Minimization of the objective function is then carried out
iteratively starting from a first guess. 
Thus, a simple physical model of the system can be used to provide a
(possibly  inconsistent) first guess which can then be made accurate by the
equations-of-motion method.

To illustrate, consider a Hamiltonian, of relevance in the nuclear collective model
\cite{Diep,TR,RT},
\be \hat H_\alpha = -\frac{1}{2M} \nabla^2 + \frac{M}{2}[(1-2\alpha) \beta^2 + \alpha
\beta^4] \label{eq:H}\ee
for an object in a five-dimensional Euclidean space;
$\beta$ is a radial coordinate in harmonic oscillator units, $\nabla^2$ 
the Laplacian, and $M$ is a dimensionless mass parameter.
Such a Hamiltonian is of general interest as a model of a system with two phases:
when $\alpha < 0.5$, the potential 
\be
V_\alpha(\beta) = \frac{M}{2}[(1-2\alpha) \beta^2 + \alpha \beta^4] 
\label{eq:pot} \ee
has a  spherical minimum (at $\beta_0 =0$); and when $\alpha > 0.5$, it has a minimum given by
$\beta_0^2 = (2\alpha -1)/2\alpha$.
It is invariant under the group of SO(5) 
rotations in the five-dimensional space. Thus, its eigenfunctions are products of $\beta$ wave
functions and SO(5) spherical harmonics.
For a state of SO(5) angular momentum $v$, the $\beta$ wave function is an
eigenfunction of the radial component of $\hat H_\alpha$
\be \hat H_\alpha^{(v)} = -\frac{1}{2M}\frac{d^2}{d\beta^2} + 
\frac{(v+1)(v+2)}{2M\beta^2}
+\frac{M}{2}[(1-2\alpha) \beta^2 + \alpha \beta^4] .  \label{eq:Hv}              
\ee

This radial Hamiltonian can be expressed \cite{Row05} in terms of an su(1,1) Lie algebra
spanned by 
\be\begin{array}{c}
\displaystyle
\hat X_1^{(v)} = \frac{d^2}{d\beta^2} - \frac{(v+1)(v+2)}{\beta^2} \,, \\ 
\hat X_2^{(v)} = \beta^2, \quad \displaystyle\hat X_3^{(v)} = 1+ 2\beta \frac{d}{d\beta} \,.
\end{array}
\ee
In terms of su(1,1) raising and lowering operators,
\be
\begin{array}{c}
\displaystyle
\hat S^{(v)}_0 = \frac{1}{4} \bigg[ -\frac{\hat X^{(v)}_1}{M} + M\hat
X^{(v)}_2\bigg],\\
\displaystyle
\hat S^{(v)}_\pm = \frac{1}{4} \bigg[ \frac{\hat X^{(v)}_1}{M} + M\hat X^{(v)}_2\mp
\hat X^{(v)}_3\bigg] 
\end{array}\ee
which satisfy the  commutation relations
\be
\big[\hat S^{(v)}_0, \hat{S}^{(v)}_\pm \big] = \pm \hat S^{(v)}_\pm ,\quad
\big[\hat S^{(v)}_-, \hat S^{(v)}_+\big] = 2 \hat S^{(v)}_0 , \quad\label{eq:su11CR}   
\ee
we obtain
\bea
\hat{H}^{(v)}_\alpha & = & 2 (1-\alpha) \hat{S}^{(v)}_0 - \alpha \left( \hat{S}^{(v)}_+ +
\hat{S}^{(v)}_- \right) \nonumber \\
& & + \frac{\alpha}{2 M} \left( 2 \hat{S}^{(v)}_0 + \hat{S}^{(v)}_+ +
\hat{S}^{(v)}_- \right)^2. \label{eq:ham}
\eea
 
For states of SO(5) angular momentum $v$, the su(1,1) Casimir invariant 
\be \hat C_{v}({\rm su(1,1)}) = \hat S_0(\hat S_0-1) - \hat S_+ \hat S_- \ee
takes the value
\be c_v = \textstyle\frac{1}{4} \big(v+\frac{5}{2}\big)\big(v+\frac{1}{2}\big) .\ee

To start the minimization process, a first guess is provided for small $\alpha$
(compared to the critical value $\alpha_c=0.5$) by the RPA that retains only quadratic terms
in a Taylor expansion of the Hamiltonian. In the present example, this amounts to dropping the
quartic term in the potential and making the approximation
\be
V_\alpha(\beta) \approx \frac{M}{2}(1-2\alpha) \beta^2  .
\label{eq:RPA} \ee
For large $\alpha$, the Hamiltonian $\hat H^{(v)}_\alpha$
can be approximated by its asymptotic
limit, obtained from a Taylor expansion of
$V_\alpha(\beta)$ about its minimum $V_\alpha (\beta_0)$,
\be \hat H^{(v)}_{\rm as} = -\frac{1}{2M}\frac{d^2}{d\beta^2} + \
\frac{(v+1)(v+2)}{2M\beta_0^2}
+\frac{M\omega^2}{2}(\beta-\beta_0)^2 .
\ee
This approximation becomes precise as $\alpha \to\infty$ or, for $\alpha > 0.5$, as
$M\to\infty$.
The physical content of these two limiting solutions is clear.
The first is that of a spherical harmonic vibrator.
The second is that of a rotor in a five-dimensional space, with moment of inertia
$M\beta_0^2$, coupled to a harmonic radial $\beta$-vibrator.
Thus, with the substitution
\bea &\displaystyle\beta-\beta_0 = \frac{1}{\sqrt{2M\omega}} (c^\dag + c) ,\\
&\displaystyle\frac{d}{d\beta} = \sqrt{\frac{M\omega}{2}}\, (c-c^\dag).\,
\eea
the asymptotic Hamiltonian is given by 
\be \hat H^{(v)}_{\rm as} = \omega (c^\dag c + \hf) +\displaystyle
\frac{(v+1)(v+2)}{2M\beta_0^2}.\ee
Corresponding approximations for the su(1,1) operators are given by
\bea &\hat X_1 = \displaystyle\frac{M\omega}{2} (c^\dag c^\dag + cc-c^\dag c -cc^\dag) -
\frac{(v\!+\!1)(v\!+\!2)}{\beta_0^2} ,&\\
&\hat X_2 = \displaystyle\big[ \beta_0 +\! \frac{1}{\sqrt{2M\omega}} (c^\dag +c)\big]
\big[\beta_0 +\! \frac{1}{\sqrt{2M\omega}} (c^\dag +c)\big]
,& \quad\\
&\hat X_3 =\displaystyle1+ \big[ \sqrt{2M\omega}\,\beta_0 +  (c^\dag
+c)\big] (c-c^\dag).&
\eea
These approximations, provide first guesses for $\alpha < 0.5-\delta$ and $\alpha > 0.5+\delta$
respectively. In the transition region, $|\alpha - 0.5| \leq \delta$, one can proceed in
steps using the results of one calculation as a first guess for the next.

\begin{figure}[htp]
  \centerline{\includegraphics[width=3.3in]{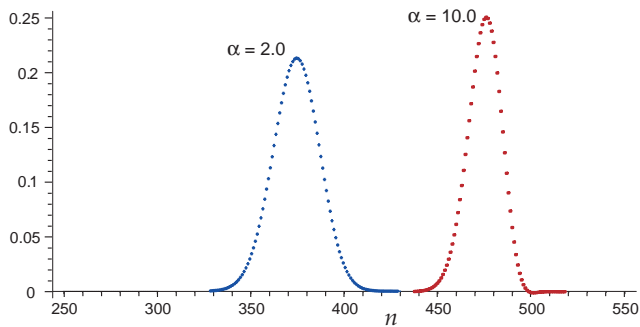}}
  \caption{Expansion coefficients for an expansion of the ground states of $\hat H(\alpha)$
   with $\alpha =2,10$ and $M=2000$  in the spherical $(\alpha =0)$ harmonic oscillator basis.
   \label{fig:coefs}} 
\end{figure}

Before presenting results, it is
instructive to consider what would be required for accurate results by diagonalization
methods.
 Fig.\ \ref{fig:coefs} shows the expansion coefficients for the $v=0$ ground state in a
harmonic oscillator basis for $\alpha = 2,10$ and $M=2000$.
The figure suggests that at least 420 basis states, for $\alpha =2.0$, and 510 basis states,
for $\alpha = 10.0$, would be needed for  the ground-state wave
function. 
More basis states would be required for excited states. 
However, because the equations-of-motion approach does not use a predetermined basis,
it is possible to obtain accurate values for energy eigenvalues for low-energy states
and matrix elements in a basis of energy eigenstates with much smaller matrices.

The number of unknowns to be determined is
reduced by exploiting the fact that $\hat S^{(v)}_0$ is self-adjoint and $(\hat
S^{(v)}_+)^{\dag} = \hat S^{(v)}_-$.    
The lowest energy level was set to zero and the energies $\{E_i\}$
of excited states regarded as unknowns.  
For the calculations reported here, the function, $F'$ was evaluated using {\it Maple} and
minimized   using  the {\it `lsqnonlin'} algorithm in {\it Matlab}.  
Results obtained with $N=10$ and a range of $\alpha$ and $M$ values are shown
below.   Minimization of $F'$ determines 197 unknowns
and approximately satisfies 500 equations.
The values of $m$ for which the entries of the $m\times m$ submatrices 
are accurate to better than 8 significant figures are listed in Table \ref{table0}.  
Numerically precise reference results were obtained by diagonalizing very large Hamiltonian
matrices in the spherical harmonic basis. 

\begin{table}[ht] 
\caption{\label{table0} 
Values of $m$ for the $m\times m$ submatrices, obtained with  $N=10$
for $v=2$, whose elements are accurate to at least 8 significant figures.
The numbers in brackets are the estimated sizes of
the Hamiltonian matrix required to  achieve similar accuracy by  diagonalization. 
}
\begin{ruledtabular}
\begin{tabular}{|c|c|c|c|c|} 
 $\alpha$ \; \; & $M = 5$ 			& $M = 50$  &  $M = 500$  &   $M = 5000$   \\ \hline
 0.2 \; \;      &  $m=$7 (25)       & 8 (15)	&  9 (20) & 9 (20)\\
 0.5 \; \; 			&  7 (30)  		& 7 (20)	&  7 (45) & 7(95) \\
 0.7 \; \; 			&  7 (30)  		& 6 (20)	&  6 (75) & 7 (490)\\
 1.0 \; \; 			&  7 (35)       & 5 (25)	&  7 (100) & 8 (770) \\ 
 2.0 \; \; 			&  7 (45)       & 5 (35)	&  7 (135) & 8 (1065) \\
 5.0 \; \; 			&  6 (65)       & 6 (60)    &  8 (175) & 8 (1235)\\
 10.0	\; \; 		&  6 (80)       & 6 (85)    &  8 (200) & 8 (1295)\\
 50.0	\; \; 		&  5 (150)      & 7 (180)  &  8 (310) & 8 ($>$ 1500)\\ 
\end{tabular}
\end{ruledtabular}
\end{table}

Typical computation times to obtain the result shown ranged from a few seconds to tens
of  seconds, achieving minimum values of $F'\sim 10^{-20}$. 
A  satisfying feature of the equations-of-motion approach is that its advantages
over conventional diagonalization are most pronounced for large values of $M$
for which  the diagonalization approach is most slowly convergent. 
Worst case scenarios for the equations-of-motion approach are when $M$ is small and $\alpha$
is large and when $M$ is large and $\alpha = 0.5$.
In the former case, the vibrational fluctuations about the equilibrium deformation are large,
and in the latter case the critical point, $\alpha = 0.5$, is highly singular.
Even though the time taken to reach a minimum in such situations may be long, the
results are invariably accurate. 
It is also noteworthy that, in the absense of good starting guesses, it is always possible to
progress in steps from previously found solutions.

For the present model, it so happens that the  $\alpha = 0.5$ results are among the easiest to
obtain for any value of $M$.
This is because of a critical point scaling symmetry \cite{RTR} which means that if the results
are known for one value of $M$ they can simply be inferred for any $M$.  
For example, the Hamiltonian at $\alpha = 0.5$ can be expressed 
\be \hat H_{0.5} = -\frac{1}{2M} \nabla^2 + \frac{M}{4} \beta^4 
= \frac{1}{2M^{1/3}}\Big[ -\bar\nabla^2 + \frac{1}{2}
\bar\beta^4\Big],\ee
with $\tilde\beta = M^{1/3} \beta$.
Thus, the energy-level spectrum of $\hat H_{0.5}$ is independent of $M$ to within an
$M^{-1/3}$ scale factor.

Table \ref{table1} gives an indication of the accuracy obtainable, for states
of interest, by equations-of-motion
calculations.
\begin{table}[t]
  \caption{\label{table1}Comparison between exact excitation energies and values calculated
  from the equations of motion for $\alpha = 2$, $v=2$, and $M=50$ for different values of
  $N$.}
  \begin{ruledtabular}
 {\scriptsize
\begin{tabular}{|c|c|c|c|} 
      Exact     & $N = 4$        & $N = 6$        &  $N = 10$       \\ \hline
 2.420379968671 & 2.420382 & 2.420379969          &  2.420379968671 \\
 4.799716618277 & 4.7999   & 4.7997167            &  4.799716618277 \\
 7.136170688911 & 7.01     & 7.136180             &  7.136170688912 \\
 9.427750466577 & -              & 9.4280         &  9.42775046660 \\ 
 11.672330365421& -              & 11.48          &  11.672330368 \\
 13.867713374832& -              & -              &  13.8677136 \\
 16.011789014165& -              & -              &  16.01180 \\
 18.102896804066& -              & -              &  18.1033 \\
 20.140619242614& -              & -              &  20.152 \\
  \end{tabular}
 }
  \end{ruledtabular}
\end{table}
It compares  excitation energies for the Hamiltonian, $\hat
H_\alpha$ with $\alpha = 2$, $v=2$, and $M=50$,  obtained for various subspace dimensions
with those of precise calculations.  Even a calculation
with $N=4$ gives accurately the first excitation energy to better
than one part in $10^6$. 
This numerical accuracy is sustained for calculated energies up to the $N-3$ level.

Table \ref{tab:2} shows the excitation energies obtained by the
equations-of-motion method with $N=5$ for the lowest two states of each SO(5) angular
momentum
$v=0,\dots ,5$ for $M=50$ and a range of values of $\alpha$.
The results are precise to better than the level of precision shown.

\begin{table}[ht]
\caption{\label{tab:2} The lowest two energy levels ($n=0$,1) with SO(5) angular
momentum
$v=0,\dots,5$, relative to the  ground state energy, computed by the equations-of-motion method
with $M=50$ and $N=5$. The results agree with those obtained by  precise calculations to
at least the level of precision shown. Results of the asymptotic approximation (AS) are
shown in the last column for $\alpha = 10$. }
\begin{ruledtabular}
\begin{tabular}{|r|c|c|c|c|c|}
$n\;\;\;\; v$ & $\alpha =0$ & $\alpha =0.5$ & $\alpha =1$ & $\alpha = 10$ & AS,\,$\alpha =10$
\\ \hline 0	\;\; 0  	&0						&0	      			&0	      		&0	      			&0       \\
0	\;\; 1	  &1						&0.40223				&0.08712			&0.04254				&0.04211 \\
0	\;\; 2	  &2						&0.84184				&0.21629			&0.10634				&0.10526 \\ 
0	\;\; 3	  &3						&1.31343				&0.38597			&0.19140				&0.18947 \\
0	\;\; 4	  &4						&1.81319				&0.59436			&0.29769				&0.29474 \\ 
0	\;\; 5	  &5						&2.33829				&0.83960			&0.42521				&0.42105 \\				\hline
1	\;\; 0	  &2						&0.96133				&1.35973			&6.13301				&6.17262 \\
1	\;\; 1	  &3						&1.45802				&1.46666			&6.17646				&6.21472 \\
1	\;\; 2	  &4						&1.97813				&1.62318			&6.24161				&6.27788 \\
1	\;\; 3	  &5						&2.52043				&1.82581			&6.32847				&6.36209 \\
1	\;\; 4	  &6						&3.08360				&2.07109			&6.43701				&6.46735 \\
1	\;\; 5	  &7						&3.66642				&2.35579			&6.56723				&6.59367 \\
\end{tabular}
\end{ruledtabular}
\end{table}

It is instructive to note that the domains in which the diagonalization and
equations-of-motion  are most successful tend to be complementary. 
Matrix diagonalization is often faster.
However, it requires much larger matrices and substantial
extra work to set up the initial matrices and interpret the results in a given basis. In
contrast, the equations-of-motion method, directly computes the matrices of all observables in
the physically relevant basis of energy eigenstates. The advantages of the equations-of-motion
are most evident for states of
$\alpha \gg 0.5$ and large $M$.  This is when states are beginning to approach their
asymptotic dynamical symmetry limits. For example, for $v=2$, $\alpha = 50$, and $M=5000$, one
can obtain all the spectroscopic properties of the $m=8$ lowest states in 93 s, using
$N=10$.  Similar accuracy was only achieved by diagonalizing larger than $1500\times
1500$ matrices and took 250 s on the same computer. If one adds in the time taken to
compute the matrices of Hamiltonian and other observables in a predefined basis,  the
equations-of-motion method wins hands down in such a situation. These advantages of the RRKK
approach are expected to be more pronounced for systems with many degrees of freedom but 
relatively simple SGA's.

An attractive property of the RRKK approach is that it makes it possible to
start from a simple approximation and make it  precise.
Thus, the RRKK equations are particularly relevant for the description of systems that
exhibit quantum phase transitions with variation of parameters, as this letter demonstrates. 
A natural possibility is to start with a mean field description of a phase
transition and, on either side of the critical point, to use the  RRKK equations of motion to
add the fluctuation contributions that are missing in the mean field treatment.

\end{document}